**Title**

# Physiology-informed layered sensing for intelligent human-exoskeleton interaction

**Authors**

Chenyu Tang[†1], Yu Zhu[†1,2], Josée Mallah[†1], Wentian Yi[1], Luyao Jin[1,3], Zibo Zhang[1], Shengbo Wang[4,5], Muzi Xu[1], Ming Shen[2], Calvin Kalun Or[6], Shuo Gao[4], Shaoping Bai[*7], and Luigi G. Occhipinti[*1]

**Affiliations**

[1]Department of Engineering, University of Cambridge, Cambridge, UK

[2]Department of Electronics Systems, Aalborg University, Aalborg, Denmark

[3]Department of Mechanical and Automation Engineering, The Chinese University of Hong Kong, Hong Kong SAR, China

[4]School of Instrumentation and Optoelectronic Engineering, Beihang University, Beijing, China

[5]Department of Electrical and Electronic Engineering, The University of Hong Kong, Hong Kong SAR, China

[6]Department of Data and System Engineering, The University of Hong Kong, Hong Kong SAR, China

[7]Department of Materials and Production, Aalborg University, Aalborg, Denmark

[†]These authors contributed equally: Chenyu Tang, Yu Zhu, Josée Mallah
[*]Correspondence to: Shaoping Bai (shb@mp.aau.dk) and Luigi G. Occhipinti (lgo23@cam.ac.uk)

**Abstract**

Wearable exoskeletons hold transformative promise for restoring mobility across diverse users with muscular weakness or other impairments. However, their translation beyond laboratory environments remains limited by sensing systems that capture movement but not underlying physiology. Here, we present a soft, lightweight smart leg sleeve that achieves anatomically aligned, layered multimodal sensing by integrating textile-based surface electromyography (sEMG) electrodes, ultrasensitive textile strain sensors, and inertial measurement units (IMUs). Each sensing modality targets a distinct physiological layer: IMUs track joint kinematics at the skeletal level, sEMG monitors muscle activation at the muscular level, and strain sensors detect skin deformation at the cutaneous level. Together, these sensors provide real-time perception to support three core objectives: controlling personalized assistance, optimizing user effort, and safeguarding against injury risks. The system is skin-conformal, mechanically compliant, and seamlessly integrated with a custom exoskeleton (<20 g total sensor and electronics weight). We demonstrate: (1) accurate ankle joint moment estimation (RMSE = 0.13 Nm/kg), (2) real-time classification of metabolic trends (accuracy = 97.1%), and (3) injury risk detection within 100 ms (recall = 0.96), all validated on unseen users using a leave-one-subject-out protocol. This work establishes a physiology-aligned sensing architecture that reframes exoskeleton perception from motion tracking to real-time physiological decoding, offering a pathway towards intelligent, adaptive, and personalized wearable robotics.

## I. Main

Wearable exoskeletons offer the potential to assist and support mobility in both healthy individuals and patients with neuromuscular impairments [1-3]. While recent advances in lightweight actuators [4, 5], ergonomic soft exosuits [6], and adaptive control strategies [7] have greatly improved mechanical and computational performance, real-world deployment remains limited. This gap arises from a structural imbalance: while exoskeleton intelligence has advanced on the actuation side, a disconnect remains on the perceptive functions linked to the wearer physiological parameters, which can only be obtained from multi-modality sensors in constant and intimate contact with skin. This imbalance constrains the system's ability to sense, model, and respond in real-time to the user's internal physiological state when it matters most [14].

Artificial intelligence (AI)-based controllers of the human-exoskeleton biomechanical system, including those based on human-in-the-loop adaptation and end-to-end deep learning, have been developed to work with signals that primarily capture the mechanical rather than the biological aspects of the wearer [8-12]. Inertial sensors, joint encoders, and plantar pressure arrays provide valuable information on kinematics, yet they capture only coarse proxies of motion intent, joint moment, and metabolic effort [6, 8, 9, 11]. As a result, even the most advanced systems remain unable to personalize assistance or ensure safety based on the user's true physiological needs [13-15]. A paradigm shift is therefore required: exoskeletons must evolve from perceiving motion to decoding physiology. As summarized in Table 1, representative systems remain single-modality and shallow in perceptual depth, excelling at phase detection or kinematic estimation but rarely accessing the layered physiological variables that govern assistance, effort, and risk.

We define three key physiological dimensions that exoskeletons must sense to operate effectively in real-world environments: *i)* joint moment as the mechanical target for assistance, *ii)* metabolic cost as the energetic objective for optimization, and *iii)* injury risk as the safety constraint requiring immediate feedback. Existing systems typically address these elements in isolation. Joint moment estimation is often hidden within non-interpretable mappings, metabolic monitoring depends on cumbersome respiratory setups, and injury detection relies on fixed force thresholds that ignore individual context. What is lacking is a compact, deployable, physiology-aligned sensing architecture capable of perceiving across the body's layers and closing appropriate feedback loops. Emerging skin-conformal sensors for strain, pressure, and muscle activity point to this direction, but they have rarely been unified into a coherent, multi-layer physiological stack for human-exoskeleton interaction [16-19].

Here we introduce a unified, physiology-layered perception architecture that functions as a form of physiological tomography that decodes the body's state layer by layer, implemented through a smart textile leg sleeve. The leg sleeve-integrated sensors suite serves as an essential companion device seamlessly integrated in smart garment through a scalable printing technique, enabling 3 fundamental tasks in the control of assistive wearable exoskeleton devices supporting: a) real-time adaptation in personalized mobility assistance, b) reduction of the overall metabolic energy consumption, and c) safeguarding the users against unwanted actions or excessive strain at the human-exoskeleton interface. The design follows a biomimetic principle along the bone-muscle-skin hierarchy: IMUs capture skeletal kinematics, textile sEMG electrodes record muscle activation, and printed graphene-based strain sensors detect skin deformation, together providing simultaneous access to kinematic, muscular, and cutaneous layers (Fig. 1a) [20-24]. The key novelty that results from the proposed system architecture lies not in the aggregation of sensors but in the structured reasoning between them. Each layer is coupled to a task-specific neural decoder that closes a distinct physiological loop, i.e. *i)* joint-moment estimation for assistance control, *ii)* metabolic-trend classification for effort optimization, and *iii)* rapid risk detection for safety assurance—within a single, lightweight platform weighing less than 20 g (Fig. 1b).

We evaluate this architecture on a custom ankle exoskeleton (Fig. 1c) using a leave-one-subject-out protocol across three representative scenarios: biomechanics-lab gait trials, outdoor assisted walking with metabolic monitoring, and controlled perturbations simulating safety-critical events. Across users and conditions, the system achieves an ankle moment RMSE of 0.13 Nm/kg, a real-time metabolic trend classification accuracy of 97.1%, and injury risk detection within 100 ms with a recall of 0.96. These results demonstrate that concurrent closure of multiple physiological loops is feasible and beneficial. Compared with the single-layer, proxy-based approaches summarized in Table 1, this work reframes exoskeleton perception as a unified, physiology-first problem, positioning sensing as the central principle, rather than a peripheral component, of human-adaptive assistance.

## II. Results

### Decoding physiology from the skin up: prototype hardware development

The smart leg sleeve integrates inertial, muscular, and cutaneous sensing layers to enable real-time, physiologically-informed interaction between humans and wearable exoskeletons (Fig. 1a). The system combines textile-based strain sensors, sEMG electrodes, and embedded IMUs into a soft, lightweight (<20 g) and anatomically aligned form factor. Each modality targets a distinct physiological layer—joint kinematics, muscle activity, and skin deformation—providing high-fidelity input for neuromechanical inference. This compact platform achieves perceptive sensing and modeling functions by *i)* decoding motion intent, *ii)* monitoring metabolic effort, and *iii)* detecting injury-related risk.

The textile strain sensors are screen-printed over the posterolateral heel region along the calcaneofibular ligament path (Fig. 1b). A structured graphene layer with ordered microcracks, formed via a single-step printing process, enables high sensitivity to localized skin deformation under small strains (<5%), with gauge factor exceeding 100 [23]. To enhance print quality and functional repeatability, we apply a sodium carboxymethyl cellulose (CMC-Na) starching treatment that ensures surface uniformity and prevents ink over-penetration into the fabric. These strain sensors are optimized to detect subtle perturbations at the human-exoskeleton interface [24].

Textile sEMG dry electrodes are fabricated by coating cotton substrates with a graphene/PEDOT:PSS composite. The porous coating provides a large effective contact area, resulting in low skin–electrode impedance comparable to wet electrodes within the sEMG bandwidth (10–500 Hz) [20, 21]. The electrodes are sewn over the tibialis anterior, fibularis brevis, and gastrocnemius muscles to capture region-specific activation patterns relevant to joint moment and metabolic dynamics.

All sensors are stitched into a commercially available compression sleeve, with conductive threads routing signals to two compact wireless readout modules (Fig. S1, Fig. S2). Each module handles one sensing modality—strain or sEMG—and includes a dedicated IMU for capturing segmental kinematics from the foot and shank, respectively. To prevent capacitive coupling, insulating encapsulation is applied to sEMG lines on the inner sleeve surface. This modular design ensures unobtrusive wearability, robust signal acquisition, and synchronized multimodal input. Detailed characterization results of the sEMG and strain sensors are provided in Table S1.

To support evaluation under controlled loading conditions, the leg sleeve is interfaced with a custom soft ankle exoskeleton (Fig. 1c, Fig. S3, Table S2). The exoskeleton comprises a back-mounted actuation unit with Bowden cable transmission links and load cells at the terminal cuffs. It enables precise delivery of joint-level torque and gait perturbations across diverse experimental conditions. This configuration provides a versatile and reproducible testbed for assessing the system-level perception performance across biomechanical, metabolic, and safety-critical scenarios.

**Three-phase experimental validation of physiology-aligned neural decoding**

To evaluate the layered sensing platform across multiple functional objectives, we designed a three-phase experimental protocol, each targeting a distinct decoding task (Fig. 2). The protocol spans (1) controlled laboratory conditions, (2) real-world overground walking, and (3) safety-critical perturbation scenarios to validate the system's ability to decode motion intent, metabolic state, and potential harm from multimodal sensor inputs.

*Task I: Joint moment estimation in a biomechanics laboratory*

To develop a decoder for estimating ankle joint moments in real time, we first conducted experiments under controlled laboratory conditions (Fig. S4) with the exoskeleton operating in a zero-torque mode. Subjects were instructed to walk at their self-selected normal pace on level ground while wearing the smart leg sleeves and soft ankle exoskeleton, with ground-truth ankle joint moments computed from motion capture and force plate data using the open source simulation software *OpenSim* inverse dynamics (Fig. 2, top left). Subject demographic details are provided in Table S3. During this phase, we verified that the exoskeleton's passive structure did not alter natural gait biomechanics: ankle angle and joint moment trajectories under no-exoskeleton and zero-torque conditions were in good agreement, with Pearson correlation coefficients exceeding 0.98 across both legs (Fig. 3a). This biomechanical consistency validates the zero-torque mode as a reliable baseline condition for decoder training.

The joint moment decoder was trained using both sEMG and IMU signals as inputs (Fig. 2, top right), leveraging the complementary strengths of the two modalities: IMUs provided phase timing and gait structure, while sEMG refined amplitude-level variations in muscle force output. Importantly, under zero-torque mode, the total joint moment arises solely from muscle contraction, referred to as biological torque. In this case, biological torque is equivalent to net joint moment, since no external mechanical assistance is applied. This equivalence provides a clean, interpretable training condition for isolating user-generated force and serves as the foundation for downstream physiological decoding tasks.

*Task II: Real-world metabolic trend monitoring under varying assistive strategies*

To investigate effort-aware exoskeleton control, we deployed a joint moment decoder during overground walking with active assistance (Fig. 3b). Unlike the zero-torque setting, where biological and net torque are equivalent, active assistance introduces additional mechanical torque from the exoskeleton actuators. In such cases, the net joint moment represents the sum of biological torque (muscle-driven) and exo torque (mechanically applied). Because IMU signals capture the resulting joint kinematics arising from both sources, they are well-suited for estimating net torque in this context. Based on this rationale, we adopted a separately trained IMU-only joint moment decoder for real-time deployment during assistive walking. This decision aligns with prior findings showing that net torque, not biological torque, is the more relevant control target under robotic augmentation [11]. Further discussion on the transition logic and torque decomposition is provided in Supplementary Note 1.

Each walking trial began with a ~3-minute zero-torque stabilization phase to allow the participant's metabolic rate to converge to a steady baseline. The exoskeleton then applied a sequence of assistive control settings, each defined by a unique pair of control law parameters: scaling factor (α) and actuation delay (Δt) (see Fig. S5, S6, and Supplementary Note 2). Each setting was maintained for approximately 3 minutes to allow the participant's metabolic response to settle into a new steady state under the given assistance condition. No rest was imposed between transitions, enabling the transitions between multiple control laws to be tested within a single continuous walking trial. For each change in control law, the transitional segment, defined as the period from the moment of parameter switch until the next metabolic steady state, was used to construct training samples that reflect the physiological impact of the new setting, including beneficial, neutral, and adverse responses (see Fig. S7).

For each assistive control transition occurring at time $t_0$, model inputs were constructed to capture the user's physiological response during the ensuing adaptation period. To ensure both input and label reflected the same transitional dynamics, we retained only those time points t for which both t - 6 s and t + 30 s remained within the same transition segment (i.e., from $t_0$ to the next steady state). This constraint avoids contamination from preceding or subsequent control laws and guarantees a causal link between the control input and the observed physiological change.

At each valid moment t, the model input was formed by concatenating a fixed 3-second context window preceding the transition ($t_0$-3 s to $t_0$) with a 6-second sliding window after the transition (t-6 s to t), using a stride of 3 s. The corresponding ground-truth label was computed as the relative change in metabolic rate between the 6-second window preceding t (t-6 s to t) and a forward-offset window centered at t + 27 s (t+24 s to t+30 s). This 30-second forward offset accounts for the physiological delay in oxygen uptake kinetics during moderate-intensity walking [25, 26], enabling each label to reflect a meaningful metabolic trend.

A relative change exceeding ±10% of the subject's zero-torque baseline was used as the classification threshold, dividing samples into increasing, decreasing, or steady effort regimes. This empirical threshold filters out minor fluctuations while retaining sensitivity to physiologically significant changes. Although metabolic adaptation is inherently delayed, the 6-second sensor input window spans approximately 4-5 gait cycles, providing a compact yet informative representation of early neuromechanical adaptation.

As illustrated in Fig. 3c, variations in α and Δt led to large inter-condition differences in metabolic rate, with certain combinations (e.g., α = 0.6, Δt = 0 ms) yielding significant energy savings, while others (e.g., α = 0.6, Δt = 350 ms) paradoxically increased energy expenditure. These outcomes reinforce the need for adaptive control schemes based on sensing feedback. Notably, these results align with prior findings that fixed, poorly tuned assistance can amplify muscular co-contraction or disrupt neuromechanical synchrony, thereby increasing metabolic cost despite torque support [27-29]. By contrast, real-time effort-aware classification provides a pathway to optimize assistive benefits on a per-user, per-condition basis. The importance of modality-specific sensing is further illustrated in Fig. 3d, where t-SNE embeddings of extracted features show that metabolic state distinctions emerge more clearly from sEMG signals than from IMU, reinforcing the critical role of physiologically grounded inputs in decoding user effort.

*Task III: Detection of biomechanical risk under sub-threshold perturbations*

The final experimental phase targeted the rapid detection of safety-critical events. In this task, subjects walked under zero-torque mode while experiencing random, low-magnitude torque pulses (<0.3 Nm/kg) applied during early (0-20%) and late (80-100%) gait cycle phases (Supplementary Note 3). These intervals correspond to foot swing and dorsiflexion, periods when only minimal force is typically needed to control ankle motion. As a result, even small and unexpected forward pushes, well below standard safety limits, can interfere with the body's sense of movement and timing, potentially causing unstable reactions such as muscle stiffening, early foot placement, or hesitation during walking.

This perturbation paradigm simulates subtle but potentially hazardous events that often occur in real-world settings, such as stepping on a small obstacle (e.g., pebble), encountering slight misalignments between ground and foot, or experiencing low-level actuator misfires in robotic systems. Although low in intensity, such events can trigger strong subjective responses due to phase-specific neuromechanical sensitivity, especially during transitions between gait subphases [30].

To capture user-perceived risk, participants carried an emergency button and were instructed to press it upon sensing instability. The corresponding strain signals at the human-exoskeleton interface exhibited abrupt temporal deviations during reported events (Fig. 3e), demonstrating the system's ability to detect biomechanical anomalies based on cutaneous-level sensing. This task highlights the additional safety value provided by skin-deformation monitoring, complementing traditional force-threshold-based approaches that may overlook subtle shifts too weak to trigger traditional safety thresholds.

## Neural decoder architecture and system-level performance evaluation

To translate raw multimodal signals into actionable control cues, we developed task-specific neural decoders for each of the three physiological objectives: joint moment estimation, metabolic trend classification, and injury risk detection. Each task selectively leverages the most relevant sensing modalities based on its physiological characteristics, rather than using all three modalities in every case. All models were trained using a leave-one-subject-out (LOSO) protocol and optimized for real-time deployment on a wearable edge processor [31].

### *Joint moment estimation decoder*

The joint moment decoder processes unilateral sEMG and IMU signals using two distinct encoder branches (Fig. 4a, left). IMU input is handled by a temporal convolutional network (TCN) with causal and dilated layers (Fig. 4c, left), chosen for its ability to model long-range sequential dependencies while preserving real-time causality—critical for motion-phase-resolved decoding [32]. sEMG input is encoded via a 1D Squeeze-and-Excitation ResNet (SE-ResNet) (Fig. 4c, right), which adaptively weights feature channels to mitigate the variability introduced by our textile sEMG electrodes. While these electrodes offer improved breathability and comfort over commercial wet sensors, they exhibit lower skin adherence. The SE module compensates for this by suppressing noisy or underperforming channels [33, 22]. The fused features are passed to a fully connected head for moment prediction. A 200 ms window with 10 ms stride was used to ensure high temporal resolution while retaining sufficient biomechanical context.

As shown in Fig. 4a (right), the model accurately tracks rapid fluctuations in joint torques. Across eight subjects, the LOSO RMSE averaged $0.133 \pm 0.015$ Nm/kg, while within-subject training achieved $0.089 \pm 0.008$ Nm/kg (Fig. 5a, top). Removing sEMG input led to modest performance degradation (LOSO: $0.159 \pm 0.017$ Nm/kg), whereas sEMG-only input performed poorly. This reflects the complementary strengths of the modalities: IMU encodes kinematic phase timing, while sEMG refines amplitude-level variations in torque. Per-subject LOSO results (Fig. 5a, bottom) show stable performance across users, all under 0.2 Nm/kg, highlighting the model's generalization capability (Fig. S8).

*Metabolic trend classifier*

The metabolic trend decoder uses bilateral sEMG and IMU signals with the same encoder structure as the moment model (Fig. 4b, left), but with adjusted kernel sizes and receptive fields to reflect the slower dynamics of metabolic changes. To capture transition context, each input sample concatenates a fixed 3 s window prior to assistive control switching with a 6 s sliding window post-switch, using a 3 s stride. This structure balances early transition sensitivity with signal richness. The model outputs a three-class label: increasing, steady, or decreasing effort. As shown in Fig. 4b (right), predicted labels closely follow ground-truth metabolic transitions triggered by assistive strategy changes.

Classification accuracy reached 98.4% within-subject and 97.1% under LOSO testing (Fig. 5b, top), with class-wise F1-scores exceeding 0.97. A context window ablation (Fig. 5b, bottom) shows that accuracy rises steeply from 2 s to 6 s, then plateaus—indicating a temporal sweet spot for metabolic trend decoding. Notably, sEMG-only input retained reasonable accuracy (~92.4%), while IMU-only performance dropped below 80%, confirming that EMG dominates metabolic inference (Fig. S9). This contrasts with joint moment estimation, where IMU carried more predictive power—highlighting task-specific modality relevance.

*Risk detection model.*

To support ultra-fast detection of biomechanical risk, we designed a binary classifier based on unilateral strain signals (Fig. 4d, top). The encoder uses a 1D ResNet trained with focal loss to emphasize rare emergency events. A 1 s input window and 50 ms stride enable rapid updates without sacrificing temporal context. Fig. 4d (bottom) illustrates the model's sub-100 ms response to user-reported emergencies, demonstrating alignment between perceived instability and signal perturbations.

Quantitatively, the model achieved 100% recall in within-subject testing and 96.4% recall under LOSO validation (Fig. 5c, top, Fig. S10). End-to-end latency was evaluated from signal acquisition on the wireless module to emergency detection on the exoskeleton's onboard processor (Jetson Nano). Most cases were classified within 100 ms of the perturbation onset, with a minority detected within the subsequent stride (50 ms later), slightly exceeding the 100 ms time length (Fig. 5c, bottom).

*System-level performance and user evaluation*

To facilitate user evaluation of system output, we developed a visual interface displaying real-time effort levels, metabolic trends, and risk alerts (Fig. 5d). We opted to present effort level as a single intuitive percentage rather than raw torque curves, which may be less interpretable for end users. The effort level is computed from sEMG signal features using classical root-mean-square extraction, not as a definitive physiological metric but as an interpretable proxy for front-end visualization, chosen for its simplicity and real-time robustness.

We further conducted a user study ($n = 12$) using a standardized Likert-scale questionnaire (Fig. 5e) designed from the crucial concept of "sense of agency" [34]. Across all items, responses were ≥4 (neutral or positive) for every participant. Particularly strong ratings were seen for comfort, reliability, and overall system integration. One item—"The system responded quickly to movement changes"—received slightly lower scores. We attribute this to the fact that the moment estimation model was trained on walking trials conducted at subjects' self-selected normal speeds, typically ranging from 1.1 to 1.5 m/s (Fig. S11), without including atypically slow or fast gait conditions, which may have limited generalization under variable-speed or non-flat gait. Expanding training to include slopes, stairs, and transitions will be a focus of future work. Nevertheless, these results underscore the system's usability, physiological transparency, and translational potential for real-world exoskeleton applications.

## III. Discussion

This study presents a human-exoskeleton interaction framework that places real-time physiological perception—rather than purely kinematic sensing—at the core of intelligent assistance. By embedding sEMG, Strain and inertial motion sensors seamlessly in smart e-textile leg sleeves the proposed system enables mapping internal human states—skeletal motion, muscular activation, and skin deformation—onto dedicated sensing modalities, hence transforming the body into a structured, high-fidelity source of control information. In addition to its lightweight, skin-wearable form factor, the layered sensing platform accurately decodes joint torque, metabolic trends, and biomechanical risk in real-world scenarios. Rather than replacing existing exoskeletons, it functions as a generalizable perceptual front-end that can retrofit existing robotic systems. For instance, torque estimation can inform low-level control laws such as impedance modulation [8], metabolic classification can guide adaptive human-in-the-loop optimization [29], and risk detection can trigger immediate safety interventions. In doing so, the framework unifies physiological inputs and control logic into a coherent architecture that enhances autonomy, safety, and personalization across both commercial and research-grade wearable robotics.

Although the system shows promising generalization and task performance, several limitations remain. First, the current validation focuses on level-ground walking at a normal pace. This does not fully reflect the complexities of real-world mobility, which involves transitions across stairs, ramps, and unexpected events. These scenarios can introduce new gait dynamics and require more robust inference models and context-aware adaptations. Second, the evaluation was conducted with healthy participants. The physiological variability observed in clinical populations, such as individuals recovering from stroke, elderly users, or patients with neurodegenerative conditions, may significantly affect signal quality, muscle patterns, and model transferability [35]. To ensure broad clinical utility, future studies should focus on expanding to these user groups. Third, while our decoding models support real-time inference and demonstrate strong cross-subject generalization, they still rely on offline training and manual parameter tuning. To overcome this limitation, future work will focus on implementing new capabilities to update the embedded perception model in real-time, with minimal calibration and without interrupting usage. For that, further investigation is required on learning strategies such as meta-learning, active sampling, and domain adaptation to enable continual, on-device personalization. These developments will play a critical role in translating the system from a physiologically-informed demonstrator to a physiology-adaptive platform for autonomous assistance.

Looking ahead, the potential impact of this work extends beyond wearable exoskeletons. As intelligent systems become increasingly embedded into the daily activities of individuals, there is growing demand for machines that can interpret internal physiological cues and not just respond to external behaviors. The sensing framework presented here provides a transferable foundation for future technologies that require embodied intelligence [36]. These may include upper-limb prosthetic devices capable of effort-aware control, occupational robotics that detect fatigue or instability in real time, and personalized rehabilitation systems that adapt based on patient progress rather than predefined schedules. Furthermore, the abstraction of multimodal sensor data into interpretable indicators such as torque intent, effort dynamics, and risk events could be used to populate real-time digital twins of human users, supporting simulation-informed interventions and long-term health tracking [37, 38]. The same sensing logic can also support the development of general-purpose somatosensory skins for robotic systems that require human-like perception [39, 40]. As the boundary between biological and artificial systems becomes increasingly porous, sensing platforms that are compact, responsive, and physiologically grounded will play a central role in shaping the next generation of intimate and symbiotic human-machine co-adaptation, potentially contributing to future human–machine co-adaptation strategies that could enhance mobility and safety [41].

## IV. Methods

### Sensor design and fabrication

The smart leg sleeve integrates three sensing modalities—sEMG, strain, and inertial sensing—each designed to target a specific physiological layer. All sensors are fabricated and embedded using garment-compatible processes to ensure comfort, compliance, and long-term durability during dynamic human-exoskeleton interaction.

*Textile sEMG electrodes*

The sEMG electrodes were fabricated by soaking porous cotton substrates with a graphene/PEDOT:PSS composite, following our previous study protocols [20, 21]. This composite provides low skin-electrode impedance within the EMG bandwidth (10-500 Hz), strong biocompatibility, and user comfort. To modulate ink penetration and enhance electrode performance, we applied a starching pretreatment using ethyl cellulose. This step improved the uniformity and adhesion of the conductive coating and reduced skin-electrode impedance, particularly in double-sided configurations. The electrodes were sewn onto the inner sleeve layer along the tibialis anterior, fibularis brevis, and gastrocnemius muscle paths to capture localized muscle activity.

*Textile strain sensors*

Strain sensors were screen-printed onto the posterolateral region of the sleeve over a knitted elastic fabric substrate. The sensing layer consisted of multilayer graphene ink, which formed aligned microcracks due to the mechanical mismatch between the graphene and the textile substrate during drying and prestretching. These ordered cracks produced high piezoresistive sensitivity, achieving a gauge factor of more than 100 within a sub-5% strain range. To ensure consistent crack formation and strong ink adhesion, a CMC-Na starching layer was applied to the sensing region before printing. More details are available in our previous studies [23, 24].

*Sensor integration*

All textile electrodes and sensors were stitched into a commercially available compression leg sleeve using conductive threads for signal routing. Care was taken to align the sensor locations with anatomical landmarks. To prevent electrical interference, insulating encapsulation layers were applied to the inner sleeve surface above sEMG lines. Inertial signals were collected using commercial 9-axis inertial measurement units (MPU-9250), which were directly integrated into the readout modules attached to the shank and foot. Sensor placement was adjusted per participant to accommodate variations in leg morphology and anatomical proportions, such as muscle belly location and joint alignment. All sensor positioning was performed under expert supervision to ensure anatomical fidelity and signal reliability.

**Wireless sensing readout integration**

To support real-time signal acquisition during dynamic locomotion, we implemented a modular, bilateral wireless readout system for the smart leg sleeve (Fig. S2). Each leg was equipped with two compact readout modules. One module acquired signals from the textile sEMG electrodes (3 channels), while the other processed strain signals (2 channels). Both modules incorporated a low-power ESP32 microcontroller and an integrated 9-axis inertial measurement unit (MPU-9250) for concurrent kinematic sensing.

Strain signals were read through voltage divider circuits following a design established in our previous work [42], enabling high-sensitivity detection of local skin deformation. sEMG signals were amplified and digitized using a dedicated analog front-end (ADS1293), communicating with the ESP32 via SPI. All sensor data were transmitted over Wi-Fi to an on-body edge processor (Jetson Nano), where task-specific neural decoders performed real-time inference. This bilateral setup captured 6-channel sEMG, 4-channel strain, and four 9-axis IMU signals in total, enabling multimodal perception of user intent, effort, and potential risk.

**Exoskeleton design and fabrication**

A portable, self-contained soft ankle exoskeleton was developed to evaluate the layered sensing platform under both passive and assistive walking conditions. The mechanical subsystem consists of two quasi-direct-drive motors (PDA08, Dr. Sci. & Tech.) mounted on a carbon fiber-reinforced back support, each connected to a pulley and Bowden cable system that delivers controlled plantarflexion torque to the user's foot. The actuation torque is transmitted to carbon fiber footplates via lightweight cables, minimizing distal mass. A 500 N load cell (DYLY08, DYSENSOR) is positioned at the interface between the cable sheath and ankle brace, enabling real-time measurement of transmitted force and torque.

All high-level control and data logging are executed on an onboard edge processor (Jetson Nano B01, Nvidia) housed within the back module. The processor communicates with embedded motor drivers over a CAN bus interface to ensure low-latency, deterministic torque regulation. A step-down converter (24 V to 5 V) powers the processor, while a 24 V, 4 Ah lithium-ion battery pack (G24B2, ASUNCELL) supports continuous operation for up to 1.2 hours under typical walking assistance profiles. This modular architecture provides a compact and robust platform for evaluating physiologically-informed exoskeleton control in real-world scenarios. The exoskeleton adopts a cascaded architecture consisting of an outer-loop non-linear admittance controller and an inner-loop PID-based velocity controller, as described in S2.

**Participants and ethical approval**

12 healthy adult participants (8 male, 4 female; age: 25.6 ± 3.78 years; height: 177.3 ± 8.79 cm; weight: 74.9 ± 12.47 kg) were recruited for this study. All individuals reported no history of musculoskeletal or neurological impairments and provided written informed consent prior to participation. Experimental tasks were distributed across subjects based on availability and physical readiness, with eight participants completing Phase I (joint moment estimation), seven completing Phase II (metabolic condition classification), and seven completing Phase III (risk detection). A detailed breakdown of participant demographics and task involvement is provided in Table S2.

All experimental procedures involving human subjects were approved by the Department of Engineering Ethics Committee at the University of Cambridge (Reference No. 639). The study was conducted in accordance with institutional guidelines and the Declaration of Helsinki.

### Experiment protocol

*a. Phase I for joint moment estimation*

This phase aimed to collect synchronized multimodal signals and ground-truth ankle joint moments for decoder development. Eight healthy adult participants completed 20 walking trials at their self-selected normal pace along an indoor level-ground walkway located within a motion capture laboratory (Fig. S4). During each trial, participants wore the smart leg sleeve and a soft ankle exoskeleton operating in zero-torque mode to avoid mechanical interference.

Whole-body kinematics were captured using a 16-camera optical motion capture system (Vicon Valkyrie VK26, Vicon, UK), and ground reaction forces were recorded via three embedded force plates (AMTI, USA) positioned along the walkway. Surface electromyography (EMG) signals were collected bilaterally from the tibialis anterior and the medial and lateral gastrocnemius muscles using textile-based electrodes, while segmental kinematics were concurrently measured by 9-axis inertial measurement units (MPU-9250) mounted on the shanks and feet.

*b. Phase II for metabolic consumption monitoring*

This phase focused on capturing real-time changes in metabolic effort under varying assistive strategies. Seven participants walked on a flat, outdoor looped pathway (392 m in total length) located immediately outside the laboratory building, while wearing the smart leg sleeve and receiving active ankle assistance from the exoskeleton.

Each subject completed three walking trials at their self-selected normal pace. Each trial began with a 3-minute zero-torque stabilization phase to establish a metabolic baseline, followed by a sequence of assistive control conditions. These assistive conditions were defined by distinct pairs of torque scaling factor ($\alpha$) and actuation delay ($\Delta t$) values. For each trial, 4-6 different assistive profiles were tested in sequence (depending on the participant's physical capacity), with each condition lasting approximately 3 minutes. No rest periods were inserted between conditions, enabling smooth transitions and sustained metabolic monitoring throughout the entire walking session.

Metabolic consumption was recorded continuously using the COSMED K5 system (COSMED, Italy), which performs indirect calorimetry to measure oxygen consumption ($VO_2$). Throughout the trials, participants wore the COSMED K5 and exoskeleton simultaneously, allowing synchronized collection of metabolic and sensor data across all assistive conditions.

*c. Phase III for potential harm detection*

This phase was designed to evaluate the system's ability to detect biomechanical risk arising from subtle gait perturbations. Seven participants walked under zero-torque conditions on the same outdoor looped pathway while wearing the smart leg sleeve and the soft ankle exoskeleton. During each walking trial, random low-magnitude plantarflexion torque pulses (<0.3 Nm/kg) were applied via the exoskeleton at either early (0–20%) or late (80–100%) phases of the gait cycle—corresponding to foot swing and dorsiflexion, respectively. These micro-perturbations simulate common destabilizing events in real-world scenarios, such as encountering small obstacles, slight terrain irregularities, or actuator malfunctions.

Each participant completed 20 walking trials. In each trial, a few random torque pulses were delivered, and the trial was terminated immediately once the participant reported a perceived instability by pressing a handheld emergency button. This design ensured that each trial captured the subject's spontaneous response to isolated, unpredictable perturbations without foreknowledge of timing or magnitude.

Gait phase timing was monitored in real time using a custom plantar pressure insole system, which estimated gait percentage based on data from an 8-channel force sensing resistor (FSR) array embedded in each insole (hardware configuration shown in Fig. S5). Strain sensors in the leg sleeve continuously recorded local skin deformation signals near the ankle joint to capture biomechanical responses to both normal walking and perturbation events.

### Signal preprocessing and feature extraction

All sensor signals, including strain, sEMG, and IMU data, were preprocessed using standard signal conditioning techniques before being used for neural network training.

Strain signals were processed using a fourth-order low-pass Butterworth filter with a cutoff frequency of 20 Hz. This filtering step was applied to remove high-frequency noise while retaining the slow deformation patterns of the skin that are informative for gait and perturbation detection. sEMG signals were filtered using a fourth-order Butterworth bandpass filter between 20 and 450 Hz to isolate physiologically relevant frequencies. IMU signals from the accelerometer, gyroscope, and magnetometer axes were denoised using a fourth-order low-pass Butterworth filter with a 20 Hz cutoff. This filtering step reduced motion-induced jitter while preserving key features of segmental kinematics.

For the visualization presented in Fig. 3d, we performed basic feature extraction to compare sEMG and IMU modalities using t-SNE. sEMG features included root-mean-square amplitude, mean absolute value, and waveform length, computed within a 200 ms window. IMU features included the mean, standard deviation, and signal magnitude area from the same window. These features were selected to capture signal intensity, variability, and temporal structure in a compact form suitable for dimensionality reduction and visualization.

For the real-time interface shown in Fig. 5d, the displayed effort level percentage was computed from sEMG signals. We calculated the root-mean-square amplitude of the sEMG envelope from the tibialis anterior and gastrocnemius channels. This value was then normalized to each subject's maximum observed amplitude during baseline walking trials. The resulting percentage offered an intuitive estimate of muscular effort, allowing users to monitor their exertion level during exoskeleton-assisted walking. This metric served as a front-end proxy for user effort, chosen for its simplicity, interpretability, and real-time responsiveness.

### Software environment and neural decoders development

All signal preprocessing was performed on a MacBook Pro equipped with an Apple M1 Max processor. Neural decoder training was conducted using Python version 3.8.13 within a Miniconda 3 environment, with model development and optimization implemented in PyTorch version 2.0.1. All training was executed on a workstation equipped with an NVIDIA RTX 4080 GPU using CUDA acceleration, ensuring efficient convergence across the three decoding tasks. Detailed training configurations, including loss curves and validation metrics, are provided in Fig. S8, Fig. S9, and Fig. S10. All data processing scripts, model architectures, and datasets will be made publicly available upon publication through a dedicated GitHub repository.

**Mobile Interface Implementation**

A custom mobile app was developed using Flutter (Dart SDK 3.3) to visualize real-time sensor decoding outputs. The app receives effort level, metabolic trend, and risk status via UDP socket from the edge processor (Jetson Nano) at 5 Hz refresh rate. Sensor inference results are formatted as lightweight JSON packets. All UI elements, including dynamic icons and percentage bars, are rendered using Flutter's native widget system to ensure responsive and cross-platform deployment (Android/iOS). The app allows users to intuitively monitor physiological states during assisted walking.

## Acknowledgments

Luigi G. Occhipinti acknowledges funding from Endoenergy Systems Limited grant No. G119004, MathWorks via the MathWorks-CUED Small Grant Programme, the Engineering and Physical Sciences Research Council (EPSRC) grant No. EP/K03099X/1 and the British Council UKIERI project Contract No. 45371261; Shaoping Bai acknowledges the financial supports of Frode V. Nyegaards and Wife's Fund via the project ALEXO and Independent Research Fund Denmark via the project VIEXO. Shuo Gao acknowledges the funding from the National Natural and Science Foundation, Grant No. 62171014.

## Author contributions

Conceptualization: CT, YZ, LGO

Methodology: CT, YZ, JM, WY, ZZ

Investigation: CT, YZ, JM, LJ, LGO

Visualization: CT, YZ, JM, WY

Funding acquisition: LGO, SB, SG

Supervision: SB, LGO

Writing – original draft: CT, YZ, JM, WY

Writing – review & editing: All authors

## Display Items

**Table 1 | Comparative overview of sensing architectures and perceptual depth in representative wearable exoskeletons.**

| Research | Sensing modality | Sensor-body interface | Physiological signal access | Decoded function | Deployment realism | Perceptual outcome |
|---|---|---|---|---|---|---|
| Slade et al. (2022) [4] | IMU | Rigid ankle modules | Kinematics | End-to-end ankle assistance adaptation | Outdoor walking (N=10) | No explicit physiological decoding |
| Lee et al. (2023) [7] | IMU, encoders | Rigid ankle modules | Kinematics | Gait-phase estimation for preference-based control | Treadmill walking (N=12) | Accurate gait phase detection |
| Molinaro et al. (2024) [8] | IMU | Rigid hip modules | Kinematics | Hip joint moment estimation | Lab testing (N=34) | Moment RMSE = 0.14 Nm/kg |
| Kim et al. (2024) [5] | IMU, load cells | Textile soft suit | Kinematics | End-to-end hip-flexion adaptation | Indoor and outdoor walking (PD patient, N=1) | No explicit physiological decoding |
| Tricomi et al. (2024) [11] | IMU | Textile shorts | Kinematics | Gait phase estimation for hip-flexion assistance | Outdoor walking (young N=12; older N=10) | Precise phase detection |
| Lee et al. (2024) [17] | EMG | Soft patches | Muscle activation | Elbow/shoulder flexion–extension decoding | Indoor lab experiments (N=5) | Decoder accuracy = 96.2% |
| **This work** | IMU, EMG, strain | Textile leg sleeves | Kinematics, muscle, and cutaneous | Ankle joint moment estimation; metabolic trend classification; risk detection | Indoor and outdoor walking (N=12) | Moment RMSE = 0.13 Nm/kg; metabolic trend accuracy = 97.1%; risk recall = 0.96 |

**Note:** All comparisons are made from the sensing perspective; assistive or control outcomes are omitted for clarity.

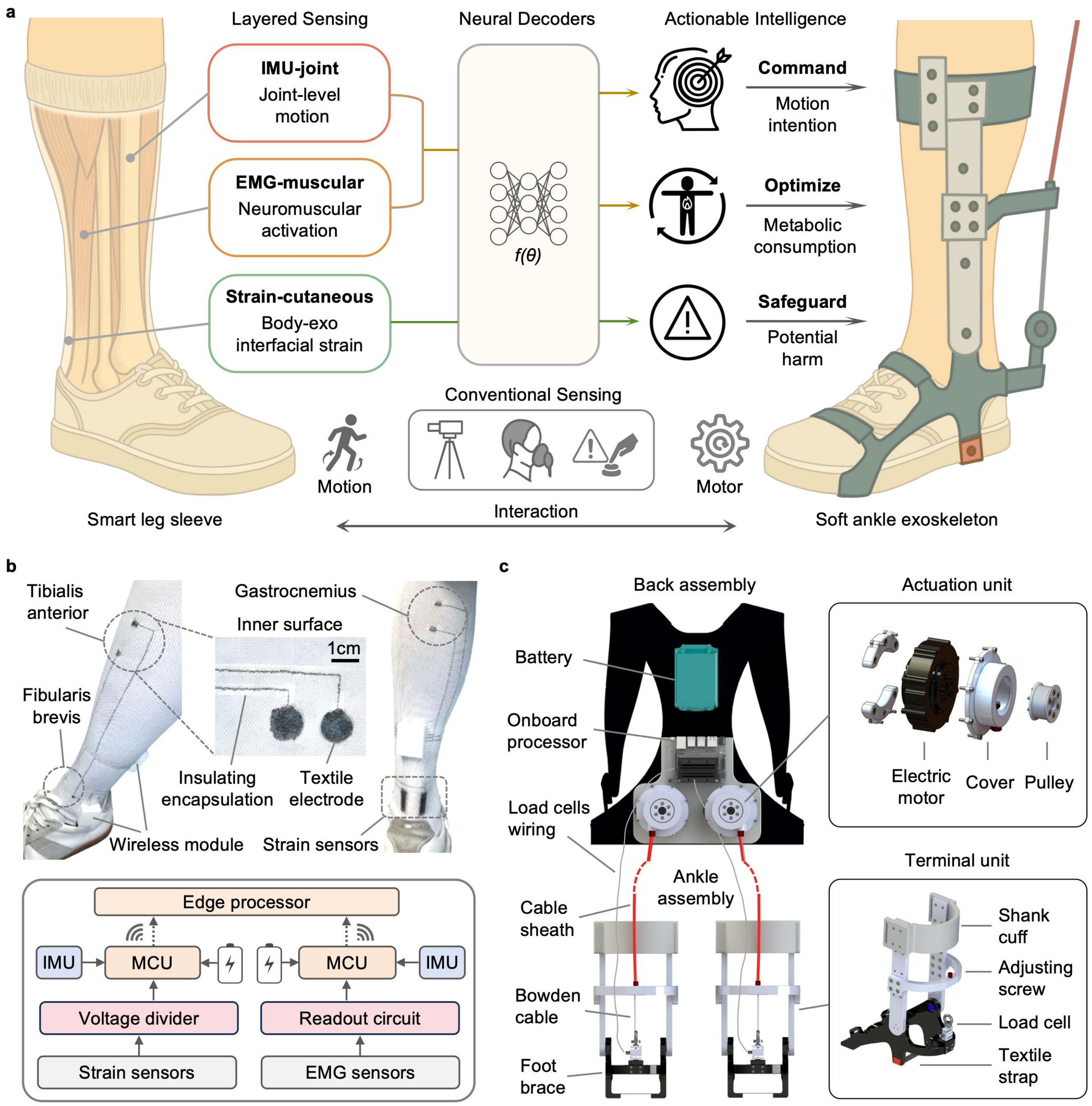

**Figure 1 | Layered smart sensing platform for physiologically informed human-exoskeleton interaction. a,** Schematic of the layered sensing architecture embedded in a smart leg sleeve, integrating inertial measurement units (IMUs), textile-based surface electromyography (sEMG), and textile strain sensors to capture neuromechanical signals at skeletal, muscular, and cutaneous levels. The decoded signals enable three core perceptual tasks: joint moment estimation for motion intent, metabolic trend monitoring for effort optimization, and injury risk detection for biomechanical safety. **b,** Hardware components and deployment of the smart leg sleeve. sEMG electrodes are placed over the fibularis brevis, tibialis anterior, and gastrocnemius to capture neuromuscular activity across the lateral, anterior, and posterior lower-leg compartments. Strain sensors are positioned bilaterally along the posterolateral heel, aligned with the calcaneofibular ligament path, to monitor skin-exoskeleton interfacial strain. IMUs are embedded in wireless readout modules for concurrent inertial and electrophysiological signal acquisition. **c,** Soft ankle exoskeleton used to validate the sensing framework. The system includes a back-mounted

actuation module with dual motors, onboard edge processor, and battery pack. Bowden cables transmit assistive torque to carbon-fibre footplates. Load cells embedded at the terminal cuffs measure the applied force during actuation.

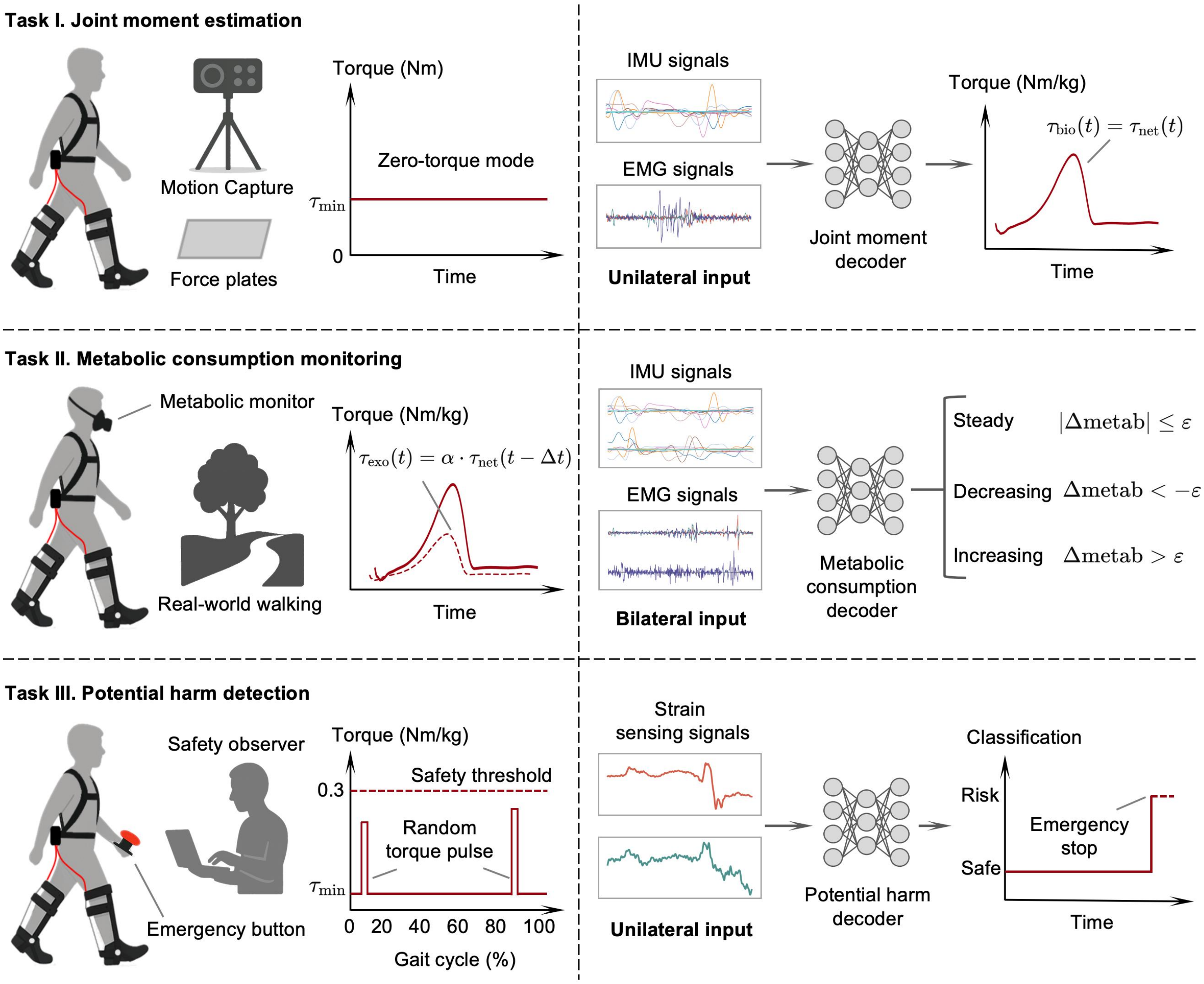

**Figure 2 | Experimental design and task-specific neural decoding pipelines for the layered sensing platform.** We designed a three-phase experimental protocol to support and validate real-time neural decoders for three core perceptual objectives: joint moment estimation, metabolic consumption monitoring, and potential harm detection. **Task I: Joint moment estimation (top).** Subjects walked in a motion capture laboratory with the exoskeleton operating in zero-torque mode. Ground-truth ankle joint moments were computed using OpenSim inverse dynamics based on motion capture and force plate data. Neural decoders were trained to estimate biological joint moment from unilateral IMU and EMG signals. **Task II: Metabolic consumption monitoring (middle).** Subjects performed overground walking in outdoor settings while receiving exoskeleton assistance modulated by varying torque scaling and delay parameters. Bilateral IMU and EMG signals were used to classify real-time metabolic trends (increasing, steady, or decreasing) inferred from paired measurements using a portable respiratory monitor. **Task III: Potential harm detection (bottom).** Under zero-torque conditions, random torque pulses within a predefined safety threshold were injected at the beginning and end of the gait cycle, mimicking scenarios where conventional force-based threshold methods fail. A decoder was trained to classify risk states from unilateral strain signals measured at the body-exoskeleton interface.

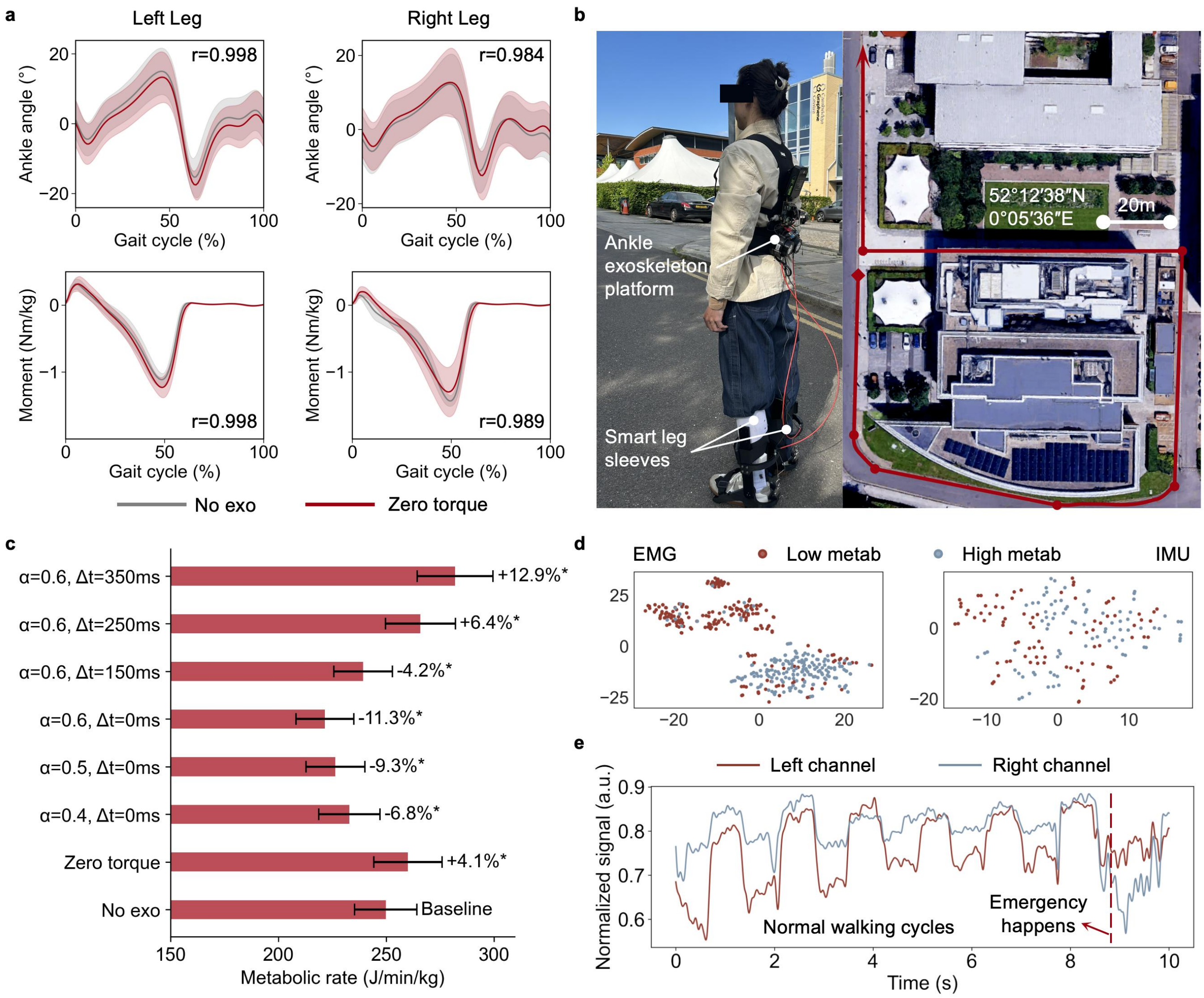

**Figure 3 | Visualization of intermediate experimental outcomes during layered sensing platform development.** **a**, Ankle angle and joint moment trajectories across the gait cycle for both legs under no-exoskeleton and zero-torque conditions. The high Pearson correlation coefficients ($r > 0.98$) indicate that the custom exoskeleton platform does not perturb natural gait biomechanics during baseline data collection. **b,** Real-world testing setup for Task II. Left: wearable configuration of the smart leg sleeves and soft ankle exoskeleton. Right: aerial view of the 392 m level-ground walking loop used for metabolic data collection. **c,** Metabolic rate under varying exoskeleton assistance parameters. Different combinations of torque scaling (α) and delay (Δt) produce significant changes in steady-state energy expenditure (*$p < 0.05$). **d,** t-SNE visualization of extracted EMG and IMU features in low vs. high metabolic conditions. Segments were labeled as high or low metabolic effort based on whether they belonged to the upper or lower 50% of each subject's steady-state metabolic values across all assistive conditions. EMG signals reveal clear metabolic state separability, while IMU features show limited discriminative power. **e,** Representative normalized strain signals during continuous walking. Sudden deviations in signal dynamics correspond to an emergency event.

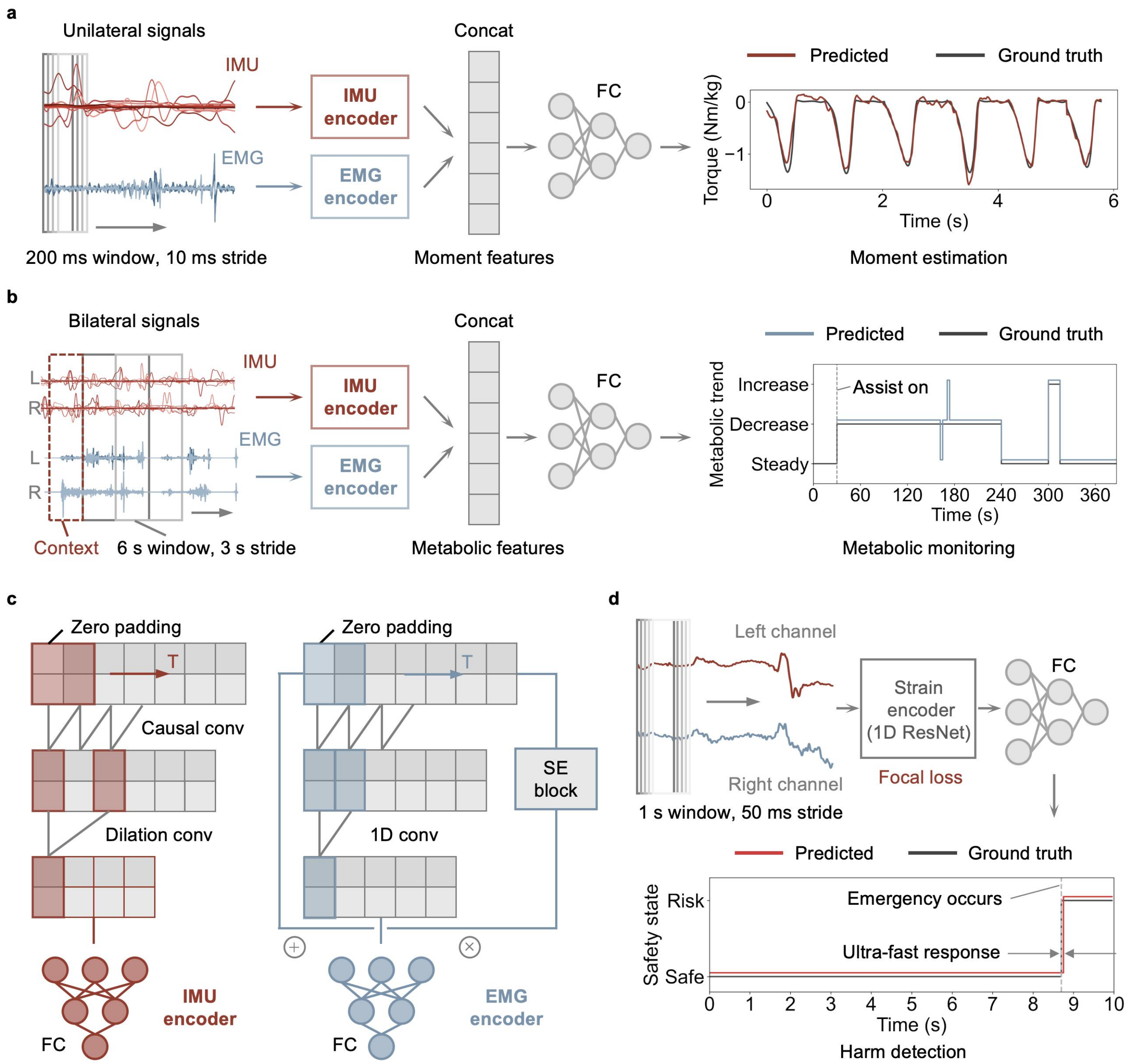

**Figure 4 | Neural decoder architectures for joint moment estimation, metabolic monitoring, and potential harm detection. a,** Pipeline for ankle joint moment decoding using unilateral IMU and EMG signals. Sensor signals are segmented into 200 ms windows with 10 ms stride and encoded via task-specific IMU and EMG encoders. The extracted features are concatenated and passed through a fully connected (FC) layer to predict continuous joint moment trajectories. **b,** Pipeline for metabolic trend classification using bilateral IMU and EMG signals. Each input segment consists of a 6 s sliding window concatenated with a fixed 3 s context segment to capture pre-transition neuromechanical state. Features extracted by the respective encoders are fused and decoded via an FC layer to classify changes in metabolic state. **c,** Architectures of IMU and EMG encoders. The IMU encoder is a temporal convolutional network (TCN) with causal and dilated convolutions for real-time processing. The EMG encoder employs a 1D SE-ResNet architecture incorporating squeeze-and-excitation blocks for channel recalibration. **d,** Pipeline for potential harm detection using unilateral strain sensing signals. A 1D ResNet encoder extracts features from 1 s windows (50 ms stride), which are decoded via an FC layer. Focal loss is employed to address class imbalance, enabling rapid and accurate detection of risk states with sub-100 ms latency.

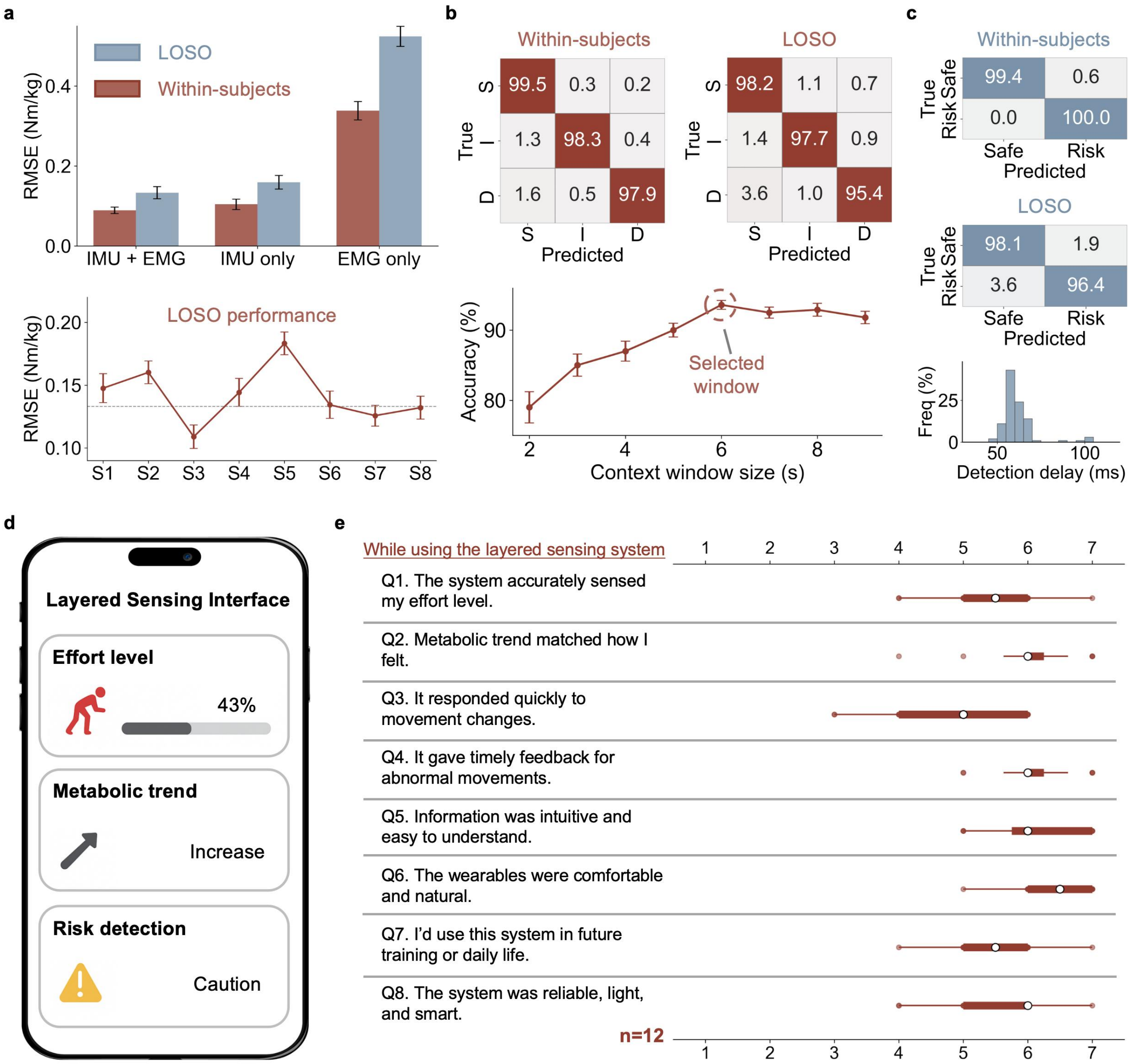

**Figure 5 | Performance evaluation of the layered sensing platform across decoding accuracy, latency, and user acceptance. a,** Joint moment estimation performance using within-subject and leave-one-subject-out (LOSO) cross-validation. The IMU+EMG combination achieved the lowest RMSE (0.08 Nm/kg within-subjects, 0.13 Nm/kg LOSO), consistently below 0.2 across all test subjects. **b,** Metabolic trend classification confusion matrices (98.6% accuracy within-subjects, 97.1% accuracy LOSO) and effect of context window size. LOSO accuracy plateaued after 6 s, reflecting the temporal integration needed for metabolic dynamics. **c,** Risk state classification via strain signals, achieving recall of 100% (within-subject) and 96.4% (LOSO). Detection delays were majorly below 100 ms. **d,** Real-time GUI displaying effort level (derived from normalized EMG envelope over gait cycle), metabolic trend, and risk alerts. **e,** User survey (n = 12) using an 8-item questionnaire rated on a 7-point Likert scale indicates high perceived accuracy, responsiveness, comfort, and usability of the layered sensing platform (1 = Strongly disagree, 7 = Strongly agree).